\documentclass[12pt,a4paper]{article}
\usepackage{amssymb,rotating,float}
\begin{document}

\title{{\bf Serine Proteases: An Ab initio Molecular Dynamics Study}}
\author{L. De Santis$^{(1,2)}$ and P. Carloni$^{(1,2,3)}$} \date{}
\maketitle

\begin{center}
$^1$ INFM -- Istituto Nazionale di Fisica della Materia\\$^2$ ISAS --
International School for Advanced Studies, \\Via Beirut 4, 34014 Trieste,
Italy \\$^3$ International Center for Genetic Engineering and Biotechnology, %
\\AREA Science Park, Padriciano 99, 34012, Trieste, Italy
\end{center}

\newpage

{\bf ABSTRACT. In serine proteases (SP's), the H-bond between His-57 and
Asp-102, and that between Gly-193 and the transition state intermediate play
a crucial role for enzymatic function. To shed light on the nature of these
interactions, we have carried out ab initio molecular dynamics simulations
on complexes representing adducts between the reaction intermediate and
elastase (one protein belonging to the SP family). Our calculations indicate
the presence of a low--barrier H-bond between His-57 and Asp-102, in
complete agreement with NMR experiments on enzyme--transition state analog
complexes~\cite{Lin_98}. Comparison with an ab initio molecular dynamics
simulation on a model of the substrate--enzyme adduct indicates that the
Gly-193--induced strong stabilization of the intermediate is accomplished by
charge/dipole interactions and not by H-bonding as previously suggested.
Inclusion of the protein electric field in the calculations does not affect
significantly the charge distribution.}

\vskip 2.0 cm


\newpage

\vspace{2cm}

\centerline{{\bf INTRODUCTION}}

Hydrogen bonding is one of the most important interactions for the
biological function of enzymes. Extensive H-bond networks may stabilize the
conformation of the active site which is necessary for the catalytic
function and they can fix the optimal orientation of the substrate for the
enzymatic reaction. Most importantly, H-bonds may allow for transition state
stabilization by lowering the activation free energy by several kcal/mol~%
\cite{Schramm_98,Fersht_85}.

A typical example in this respect is constituted by serine proteases (SP's)
enzyme family\cite
{Fersht_85,Kraut_77,Stroud_74,Branden_99,Matheson_91,Steitz_82,Blow_69,Matthews_67,Lin_98II,Cassidy_97,Frey_94}%
.

SP's use the catalytic triad (Ser-195--His-57--Asp-102) to catalyze the
hydrolysis of peptides (Fig. 1a). This occurs through nucleophilic addition
of the 3-hydroxyl group of Ser-195 to the acyl carbonyl of the substrate,
with formation of a negatively charged tetrahedral intermediate (Fig. 1b).

Stabilization of the intermediate is achieved by formation of two H-bond
with the amide groups of Ser-195 and Gly-193 (mammalian isoenzymes\cite
{Kraut_77}) or with the amide groups of Ser-195 and the sidechain of Asn-155
(bacterial isoenzymes\cite{Bryan_86}).

Theoretical\cite{Hwang_87,Warshel_89} and experimental\cite
{Bryan_86,Wells_86} studies on wild type and mutants of a bacterial SP
(subtilisin) have shown that Asn-155 is a key residue for the biological
function, in that it provides a stabilization of the transition state (TS)
relative to the ground state (GS) by as much as $\approx $ 5 kcal/mol.
Curiously, no correspondent studies on the mammalian isoenzymes have
appeared to shed light on the crucial role of Gly-193\cite{notaser}.

A second, important H-bond interaction involve two residues of the
catalytic triad, His-57 and Asp-102. A series of NMR studies on a
mammalian\cite {Lin_98,Lin_98II,Cassidy_97,Frey_94} and
bacterial\cite{Halkides_96} SP's and their complexes with inhibitors
have indicated the presence of a low--barrier hydrogen bond (LBHB)
linking N$_{\delta 1}$ of protonated His-57 with the $\beta$ carboxyl
group of Asp-102 (Fig. 1b)\cite
{Lin_98,Lin_98II,Cassidy_97,Frey_94}. Approaching of the TS is
suggested to facilitate the formation of the LBHB, which in turn may
render N$_{\epsilon 2}$ of His-57 a stronger base for accepting a
proton from Ser-195 in the formation of the
intermediate\cite{Lin_98,Lin_98II,Cassidy_97,Frey_94}. As result of
this process, the free energy barrier of TS relative to GS
decreases. Ab initio calculations and neutron scattering experiments
have led to the conclusion that the LBHB is very covalent in
nature\cite {Schiott_98}. Thus, this ``resonance--stabilization''
energy could supply much of the energy necessary for enzyme's
catalysis\cite {Lin_98,Lin_98II,Cassidy_97,Frey_94}. However, the role
of this LBHB for the catalytic power of SP's is object of
controversy\cite{Warshel_95,Warshel_98}.

From the above considerations it is clear that, in spite of its crucial
role, the nature and the dynamics of hydrogen bonding in active site of SP
are not fully understood. In order to provide a picture of the chemical
bonding of these interactions in SP's, and to relate it to the biological
function, we have carried out ab initio molecular dynamics simulations~\cite
{Car_85} on models of SP--intermediate and SP--substrate complexes. This
technique is revealing itself as a useful computational tool for
investigating specific molecular interactions in biological systems~\cite
{CP_Generali}. Based on state--of--the art density functional theory
calculations, it describes interactions and charge distributions of
(relatively) large systems rather accurately. Furthermore, it includes
temperature effects, which obviously play a fundamental role in SP function.
Finally, they allow describing bond breaking and bond forming processes at
room temperature, which is essential to describe phenomena such as the LBHB.

Anticipating our results, the calculations show that the LBHB is a strongly
covalent interaction and that the dramatic Gly-193--induced stabilization of
the reaction intermediate is due mainly to the electrostatic interactions
between the intermediate and the Asn-192--Gly-193 peptide's unit dipole.

\vspace{2cm}

\centerline{ {\bf COMPUTATIONAL PROCEDURE} }

\noindent{\bf Model complexes.}

Our structural models for the adducts of SP with intermediate (I$\cdot$SP)
and substrate (S$\cdot$SP) are based on the X-ray structure of porcine
pancreatic elastase (240 residues) complexed with
Ace-Ala-Pro-Val-difluoro-N-phenylethylacetamide (PDB entry: 4EST)~\cite
{Takahashi_89,PDB}. They include the entire catalytic triad, the scissile
peptide bond and the oxyanion hole.

The construction of the complexes is carried out in several steps: {\it i) }
the terminal N-phenylethylacetamide is replaced by an acetyl group; {\it ii)}
all hydrogen atoms of the complex, not present in the X-ray structure, are
added assuming standard bond lengths and bond angles; {\it \ iii)} a shell
of 1453 water molecules including the crystallographic ones
(approximatively corresponding to three water monolayers), is added; {\it %
iv)} four chlorine counter ions are added to ensure neutrality; {\it v)}
energy minimization is carried out with the AMBER suite of programs~\cite
{AMBER} using the AMBER force field~\cite{AMBER_FF} (convergence criterion
0.0001 kcal/(mol$\cdot $\AA )). In the minimization, no periodic boundary
condition are applied and the electrostatic interactions are calculated
assuming a constant dielectric function $\epsilon =1$ and without cutoff.

The I$\cdot$SP and S$\cdot$SP model complexes comprises the entire
side-chain of Asp-102, the imidazole ring of His-57, the Q192G193 peptide
linkage and the entire Ser-195 residue, which in I$\cdot$SP is covalently
bound to the substrate (Fig.~\ref{f:f3}a,b). His-57 is considered doubly
protonated in I$\cdot$SP and protonated in the $\delta$ position in S$\cdot$%
SP~\cite{Babine_97}.

Alternative models representing I$\cdot$SP and S$\cdot$SP differ from
the previous ones for the substitution of the Q192G193 peptide link
with dimethylammonia (Fig.~\ref{f:f3}c,d).

\vspace{1cm}

\noindent{\bf Calculations.}

The quantum--mechanical problem is solved within the framework of density
functional theory (DFT)\cite{HK}, in its Kohn and Sham (KS)\cite{KS}
formulation. The KS orbitals are expanded in a plane--wave basis set, up to
an energy cut--off of $70$ Ry. Only valence electrons are considered
explicitly, pseudopotentials of the Martins and Troullier~\cite{Troullier_91}
being used for the core--valence electron interaction. BLYP\cite
{Becke_88,Lee_88} gradient--corrected exchange--correlation functionals are
used. The charge of all the complexes is -1.

We have carried out geometry optimization using the direct inversion in
iterative subspace (DIIS) method~\cite{Pulay_80,Hutter_94} for both
electronic (convergence threshold set to $10^{-5}$) and ionic degrees of
freedom (convergence threshold set to $5\cdot10^{-4}$).

DFT--based Car--Parrinello ab initio molecular dynamics simulations~\cite
{Car_85} are performed at constant temperature, with the atomic trajectories
collected over a period of 0.6 and 1.0 ps for S$\cdot$SP and I$\cdot$SP,
respectively. Equations of motion are integrated with the velocity Verlet
algorithm. The fictitious electron mass is 400 a.u. and the integration
timestep is 4 a.u. Constant temperature simulations are achieved by coupling
the systems to a Nos\'{e}~ thermostat\cite{Nose_81} at T = 300 K with a
frequency of 500 cm$^{-1}$. The terminal hydrogens of Asp-102, His-57, and
Gly-193, corresponding to the C$_\alpha $, C$_\beta $ and C$_\alpha $
respectively, are kept fixed in their starting position; in S$\cdot$SP an
additional constraint between the O$_\gamma $(Ser-195) and the substrate
carbon of the scissile bond (indicated in Fig.~\ref{f:f3}b as C$_S$) is
imposed.

Calculations including the external electrostatic potential of the whole
protein--water system are also carried out. This potential $\Phi _{prot}(r)$
at the point $r$ is evaluated as 
\begin{equation}
\Phi _{prot}(r)=\sum_i\frac{q_i}{|r_i-r|}  \label{e:classical}
\end{equation}
where $q_i$ are the RESP~\cite{Cornell_93} atomic point charges at point $%
r_i $.

All Car--Parrinello calculations are performed with a parallel version of
the CPMD code V3.0h~\cite{CPMD_3.0h}.

\vspace{1cm}

\noindent{\bf Calculated properties.}

The electrostatic energy $\Delta E$ between to moieties (e.g. the
intermediate and Q192G193 peptide unit) is calculated as 
\begin{equation}
\Delta E=\sum_{ij}\frac{q_iq_j}{r_{ij}},  \label{e:DE}
\end{equation}
where the indexes $i$ and $j$ refer to atoms of the two moieties. 
$q_i$ and $q_j$ are the partial atomic ESP
charges\cite{Cox_82} and $r_{ij}$ is the interatomic distance.

Test calculations are carried out also using the multipolar expansion of the
electrostatic energy up to the dipolar term:

\begin{equation}
\Delta E \simeq Q_1 \Phi_2 - \mbox{\boldmath{$\mu_1$}} \cdot {\bf E_2}.
\label{e:dipole}
\end{equation}

where Q$_1$ and $\mbox{\boldmath{$\mu_1$}}$ are charge and dipole moments of
moiety 1 and $\Phi _2$ and ${\bf E_2}$ the electric potential and the
electric field produced by moiety 2. respectively. The results turn out to
be very similar to those obtained with eq. \ref{e:DE}.

Binding energies (B.E.'s) are calculated as total energies differences
between complexes in Fig.~\ref{f:f3} and their forming elements. The B.E. of
complexes {\bf I} and {\bf III} could not be determined because of the
instability of the intermediate fragment.

\vspace{2cm}

\centerline{ {\bf RESULTS} }

In this section, first we analyze structural and electronic features of two
models representing the adduct between serine protease and the reaction
intermediate ({\bf I} and {\bf III} of Fig.~\ref{f:f3}). Comparison is then
made with features of models of the substrate--enzyme complex ({\bf II} and 
{\bf IV} of Fig.~\ref{f:f3}).

\vspace{1cm}

\noindent{\ {\bf The Intermediate--Enzyme complex} }

{\it Structural features and charge distribution}. Conformational properties
as well as the H-bond network of the complex are fairly maintained during
the dynamics (Fig.~\ref{f:f4}). Consequently, the charge distribution does
not change significantly (with exception of C$_S$ and N$_S$), as indicated by the ESP atomic partial charges
reported for several snapshots of the molecular dynamics (Tab. 1). Note that
the C$_S$--O bond of the intermediate is very polarized towards the oxygen,
consistently with the fact that this bond is to be broken in the subsequent
step of the hydrolysis. The presence of the protein field does not affect
significantly the charge distribution (Tab. 2), suggesting that solvent
effects do not play a major role for the electrostatic interaction at the
active site.

{\it H-bond pattern: Asp-102--His-57.} During the dynamics, proton hopping
occurs between the His-57 and Asp-102 in the subps time--scale (Fig.~\ref
{f:f5}). The presence of a LBHB is completely consistent with NMR data on an
intermediate--serine protease complex, namely the peptidyl trifluoromethyl
ketone--chymotrypsin adduct~\cite{Lin_98}.

The chemical bonding of the LBHB can be characterized with the electron
localization function (ELF)~\cite{Becke_90,Silvi_94,Savin_97}. The ELF has
proven to be very useful to illustrate chemical concepts like localized
bonds and electron lone pairs. Fig.~\ref{f:f6} shows the ELF before, during
and after the proton transfer from His-57 to Asp-102. The red areas indicate
strong localization, i.e. spatial regions where the Pauli principle has
little influence on the electron distribution.

Fig.~\ref{f:f6}a shows the presence of the lone pairs of the aspartate
oxygen and of the strong electron localization along the histidine N$%
_{\delta 1}$--H bond, which indicates the covalent nature of the bond.
During the proton transfer (Fig. ~\ref{f:f6}b), the N$_{\delta 1}$--H bond
is still very covalent and an incipient covalent O$_{\delta 2}$--H bond is
being formed. Protonation of Asp-102 establishes a covalent O--H bond:
significant portion of ELF is indeed localized on the H atom (Fig.~\ref{f:f6}%
c). The formation of the nitrogen electron lone--pair is also evident from
the picture. We conclude that the bonding in this LBHB is essentially
covalent in nature. Similar findings have been reported in a very recent ab
initio study of low--barrier H-bonds in an organic molecule~\cite{Schiott_98}%
.

{\it H-bond pattern: Gly-193--intermediate}. The second fundamental H-bond
interaction investigated here involves Gly-193 and the intermediate carbonyl
oxygen. This H-bond is well maintained during the dynamics (average O$\cdot
\cdot \cdot $H distance of 1.7(0.1) \AA ). A rough estimation of the
interaction energy based on an electrostatic model (see Computational
Section) indicate that Gly--193 stabilizes the intermediate by more than 10
kcal/mol (Tab. 3). This value appears to be too large for a purely
electrostatic H-bond interaction~\cite{Rao_87,Jeffrey_91}.

Inspection of the structure reveals that the very large Q192G193 peptide's
unit dipole ($\approx $~4 D~\cite{Branden_99}) could be also an important
factor for intermediate stabilization, as it points towards the negative
charge of the intermediate. To extract the peptide dipolar contribution from
the total stabilization energy we construct a second model complex in which
the Q192G193 peptide unit is substituted by dimethylammonia ({\bf III} in
Fig.~\ref{f:f3}c). Tab. 3 shows that the resulting stabilization is much
smaller, only few kcal/mol. Thus, we conclude that a large contribution of
the transition state stabilization is due to electrostatic interaction
(charge--dipole interactions).

To study the relevance of the Q192G193 dipole on the dynamics, an ab
initio molecular dynamics simulation on complex {\bf III}, where the
Q192G193 peptide unit is replaced by a dimethylammonia, is performed. Fig.~%
\ref{f:f7}, which reports structural properties of the complex, indicate
that this complex is very unstable with respect to the substrate--enzyme
complex. Indeed, while the key Gly-193--intermediate H-bond becomes very
weak(Fig.~\ref{f:f7}a), the protonated His-57 transfers a proton to the
intermediate (Fig.~\ref{f:f7}a) and the O$_\gamma($Ser-195)--C$_S$ bond of
the intermediate breaks. As a result, a double C$_S$--O(I) bond is formed
(as indicated by the decrease of the bond distance up to the typical value
of a carbonyl peptide bond (1.25 \AA\ in Fig.~\ref{f:f7}b)) and C$_S$
changes it hybridization from $sp^3$ to $sp^2$, with formation of the planar
peptide unit (as shown by the increase of the $\angle $(N(I)--C$_S$--O(I))
angle up to $\approx$ 120$^o$ (Fig.~\ref{f:f7}c)). In conclusion, our
calculations suggest that the absence of the stabilizing Q192G193 dipole
causes the reverse of the reaction, with formation of the substrate and the
original H-bond pattern of the catalytic triad.

\vspace{1cm}

\noindent{\bf The Substrate--enzyme complex}

{\it Gln-192--substrate interactions}. To estimate the stabilization of the
Q192G193 peptide unit's dipole on the substrate, we perform an ab initio
molecular dynamics simulation of a model of the enzyme--substrate adduct (%
{\bf II} in Fig.~\ref{f:f3}b).

Fig. 7 shows that during the dynamics the two key H-bond interactions are
maintained but no proton transfer occurs. Interestingly, the
substrate--protein interaction energy turns out to be much lower than that
of the I$\cdot$SP complex (Tab. 3). Replacing the Q192G19 peptide with
dimethylammonia (complex {\bf IV}) causes a drastic decrease of the
interaction energy. The latter turns out to be practically identical to that
of complex {\bf III} (Tab. 3). We conclude that the H-bond interaction are
similar in the S$\cdot$SP and I$\cdot$SP complexes. In contrast, the
electrostatic (charge--dipole) interactions are very different, the I$\cdot$%
SP being more stable by $\approx$ 6 kcal/mol than S$\cdot$SP (Tab. 3). It is
interesting to note that this value compares well with previous quantum
mechanical calculations for the Asn-155--TS stabilization in the bacterial
isoenzyme\cite{Hwang_87,Warshel_89}. We conclude that the transition
state stabilization is due mostly to charge--dipole interactions.

For these complexes it has been possible to calculate also the binding
energies. Tab. 3 shows a qualitative agreement between binding energies and
energies based on electrostatic model. This validates the use of the
electrostatic model for a qualitative analysis of intermolecular
interactions, as it has been done in this work. However, it must be
stressed that use of more realistic quantum mechanical models, 
which comprises other
aminoacids present in the active site cavity, is expected to screen the
charges and therefore to reduce the calculated binding energies.

{\it Charge Distribution}. Also in this complex, most of the ESP charges do
not vary significantly during the dynamics and by introducing the electric
field of the protein (Tabs. 4 and 5). Most of the ESP charges turn out to be
similar to those of the I$\cdot$SP complex. A notable exception is
represented by the C--O peptide bond, which in this case is much less
polarized toward the oxygen. Thus it appears that the protein active site,
and in particular the Q192G193 moiety, is engineered so as to render the
scissile bond more polar in the formation of the transition state.

\vspace{2cm}

\centerline{ {\bf DISCUSSION} }

\vspace{2cm}

Within the very short time--scale here explored, our ab initio molecular
dynamics simulations help elucidate important aspects of two key
interactions in serine proteases--reaction intermediate complexes, the
His-57--Asp-102 LBHB and the Gly-193--intermediate H-bonds.

Our calculations are completely consistent with and confirm the existence of
a LBHB between His-57 and Asp-102, which has been observed experimentally in
transition state analog inhibitor complexes\cite
{Lin_98,Lin_98II,Cassidy_97,Frey_94}. Furthermore, they strongly support the
proposal of an LBHB--facilitated mechanism\cite{Lin_98}. Indeed, the LBHB
turns out to be mostly covalent in nature. The energy supplied by covalent
interaction may be crucial to overcome the energy loss due to the
compression of the two residues, which is a prerequisite of the postulated
LBHB--based reaction\cite{Lin_98}.

The second conclusion of this paper is that the rather large,
Gly-193--induced stabilization of the transition state with respect to the
fundamental state is not caused by an H-bond with Gly-193, as commonly
proposed\cite{Kraut_77,Stroud_74}: indeed, as the H-bond favors the binding
of both substrate and intermediate by $\approx $ 2.6 kcal/mol, a value
typical of a strong H-bonds in biological systems~\cite{Jeffrey_91}.
Instead, the negatively charged transition state turns out to be more stable
relative to S$\cdot $SP by several kcal/mol as a result of the interaction
of the negative charge with the large dipole of the Q192G193 peptide unit. A
simulation in which dimethylammonia replaces the Q192G193 peptide unit
confirms the crucial role of the dipole: the absence of the stabilizing
charge--dipole interaction renders the intermediate species unstable. These
considerations suggest that site--directed mutagenesis experiments on the
192 and/or 193 positions might affect significantly the activity of SP's, as
the Q192G193 dipole orientation may be not optimal for transition state
stabilization.

Because environment effects may be very important for the chemistry of the
active site of this and other enzymes~\cite
{Hwang_87,Warshel_89,Rao_87,Warshel_86}, we carry out some of the
calculations in the presence of the electric field of the protein. Our
results, summarized by tables 1--2 and 4--5, indicate however that the field
appears not to affect dramatically the charge distribution of the I$\cdot$SP
and S$\cdot$SP complexes. More sophisticated models of the protein electric
field, which for instance include the electronic polarizabilities of the
protein atoms, are not expected to alter significantly the picture.

\vspace{2cm}

\centerline{ {\bf ACKNOWLEDGMENTS} }

Vincent Torre and Frank Alber are gratefully acknowledged for their valuable
comments on the manuscript. We acknowledge financial support by
Cofinanziamento M. U. R. S. T. (Ministero dell'Universit\`{a} e della
Ricerca Scientifica e Tecnologica).

\newpage

\begin{figure}[!h]
\caption{ Schematic views of the H-bond network in mammalian serine
proteases active site (a) and of the adduct with the intermediate of the
enzymatic reaction (b). In (b) the double arrow symbol refers to the a
low--barrier H-bond.}
\label{f:f1}
\end{figure}
\begin{figure}[!h]
\caption{Model complexes representing I$\cdot$SP ((a) and (c)), S$\cdot$SP
((b) and (d)). In (c) and (d) the Q192G193 peptide unit is replaced by
dimethylammonia. H-bonds are depicted with dashed lines. Green arrows
indicate the scissile carbon atom C$_S$. The latter is labeled only in (b)
for clarity.}
\label{f:f3}
\end{figure}
\begin{figure}[!h]
\caption{Molecular dynamics of I$\cdot$SP: final structure of model {\bf I}.
H-bonds are represented with dashed lines.}
\label{f:f4}
\end{figure}
\begin{figure}[!h]
\caption{His-57--Asp-102 H-bond in I$\cdot$SP (complex {\bf I}): H--O$%
_{\delta 2}$(Asp-102) (red line) and H--N$_{\delta 1}$(His-57) (blue line)
distances plotted as a function of time.}
\label{f:f5}
\end{figure}
\begin{figure}[!h]
\caption{His-57--Asp-102 proton transfer: electron localization function
(ELF) of three snapshot during the dynamics. The ELF is represented in a
best--fit plane containing the oxygen, the proton and the imidazole ring,
and it ranges from 0 (blue) to 1 (red).}
\label{f:f6}
\end{figure}
\begin{figure}[!h]
\caption{Molecular dynamics of I$\cdot$SP: selected structural properties of
Complex {\bf III} plotted as a function of time. (a) H(Gly-193)--O(I) (blue
line), H$_{\epsilon 2}$(His-57)--O$_\gamma$(Ser-195) (green line) distances;
(b) C$_S$--O(I) bond length; (c) N(I)--C$_S$--O(I) angle.}
\label{f:f7}
\end{figure}
\begin{figure}[!h]
\caption{H-bonding of S$\cdot$SP (Complex {\bf II)}: H(Gly-193)--O(S) (blue
line), O$_{\delta 2}$(Asp-102)--H$_{\delta 1}$(His-57) (red line) distances
plotted as a function of time. }
\label{f:f8}
\end{figure}

\begin{sidewaystable}
\centerline{ {\bf TABLE 1.  I$\cdot$SP {\it in vacuo}}}
\begin{center}
\begin{tabular}[!H]{rccccccccccc}
Time~(ps) & O$_\gamma$(S195) & C$_S$ & N$_S$ & O$_S$ & N~(G193) & C~(Q192)
& O~(Q192) & N$_{\delta 1}$(H57) & N$_{\epsilon 2}$(H57) & H$_{\delta 1}$(H57) & O$_{\delta 2}$~(D102) \\ 
\hline
Init. & -0.511 & 1.194 & -0.802 & -0.717 & -0.603 & 0.504 & -0.631 & -0.009
& -0.072 & 0.303 & -0.907 \\ 
0.1 & -0.509 & 1.202 & -0.529 & -0.837 & -0.733 & 0.597 & -0.649 & -0.084 & 
~0.011 & 0.275 & -0.770 \\ 
0.2 & -0.461 & 1.022 & -0.237 & -0.724 & -0.646 & 0.571 & -0.630 & -0.171 & 
~0.058 & 0.286 & -0.688 \\ 
0.3 & -0.328 & 0.835 & ~0.006 & -0.658 & -0.556 & 0.516 & -0.631 & -0.178 & 
~0.141 & 0.298 & -0.676 \\ 
0.4 & -0.342 & 1.083 & -0.193 & -0.754 & -0.582 & 0.507 & -0.625 & -0.078 & 
-0.046 & 0.265 & -0.687 \\ 
0.5 & -0.372 & 1.146 & -0.201 & -0.794 & -0.558 & 0.505 & -0.616 & -0.025 & 
-0.150 & 0.257 & -0.665 \\ 
0.6 & -0.407 & 1.233 & -0.480 & -0.829 & -0.615 & 0.578 & -0.654 & -0.042 & 
~0.018 & 0.295 & -0.693 \\ 
0.7 & -0.430 & 1.174 & -0.311 & -0.883 & -0.570 & 0.563 & -0.638 & -0.061 & 
~0.089 & 0.284 & -0.655 \\ 
0.8 & -0.551 & 1.245 & -0.401 & -0.843 & -0.487 & 0.478 & -0.620 & -0.141 & 
~0.001 & 0.261 & -0.601 \\ 
0.9 & -0.504 & 1.124 & -0.202 & -0.772 & -0.503 & 0.502 & -0.634 & -0.127 & 
~0.070 & 0.294 & -0.591 \\ 
1.0 & -0.399 & 1.074 & -0.245 & -0.827 & -0.593 & 0.636 & -0.694 & -0.079 & 
~0.079 & 0.294 & -0.645 \\ \hline
Average & -0.438 & 1.121 & -0.327 & -0.785 & -0.586 & 0.542 & -0.638 & -0.090
& ~0.018 & 0.283 & -0.689 \\ 
St. Dev. & ~0.072 & 0.112 & ~0.207 & ~0.064 & ~0.064 & 0.048 & ~0.021 & 
~0.054 & ~0.079 & 0.015 & ~0.083 \\
\hline
\end{tabular}
\end{center}
\caption{Selected ESP partial atomic charges of I$\cdot$SP {\it in
vacuo}.}

\end{sidewaystable}

\begin{sidewaystable}
\centerline{ {\bf TABLE 2.  I$\cdot$SP with protein external field}}
\begin{center}
\begin{tabular}[!H]{rccccccccccc}
Time~(ps) & O$_\gamma$(S195) & C$_S$ & N$_S$ & O$_S$ & N~(G193) & C~(Q192)
& O~(Q192) & N$_{\delta 1}$(H57) & N$_{\epsilon 2}$(H57) & H$_{\delta 1}$(H57) & O$_{\delta 2}$(D102)
\\ \hline
0.1 & -0.513 & 1.154 & -0.531 & -0.876 & -0.646 & 0.640 & -0.750 & -0.118 & 
~0.094 & 0.354 & -0.984 \\ 
0.2 & -0.453 & 0.931 & -0.261 & -0.677 & -0.527 & 0.595 & -0.722 & -0.256 & 
~0.122 & 0.404 & -0.914 \\ 
0.3 & -0.329 & 0.649 & -0.037 & -0.580 & -0.410 & 0.481 & -0.692 & -0.273 & 
~0.147 & 0.415 & -0.890 \\ 
0.4 & -0.381 & 0.961 & -0.036 & -0.739 & -0.430 & 0.400 & -0.644 & -0.259 & 
~0.088 & 0.374 & -0.874 \\ 
0.5 & -0.412 & 1.000 & -0.106 & -0.773 & -0.358 & 0.341 & -0.611 & -0.256 & 
-0.006 & 0.397 & -0.870 \\ 
0.6 & -0.482 & 1.270 & -0.305 & -0.929 & -0.460 & 0.447 & -0.657 & ~0.015 & 
~0.100 & 0.271 & -0.804 \\ 
0.7 & -0.439 & 1.267 & -0.295 & -0.932 & -0.354 & 0.504 & -0.714 & -0.158 & 
~0.123 & 0.336 & -0.761 \\ 
0.8 & -0.602 & 1.373 & -0.382 & -0.843 & -0.349 & 0.462 & -0.720 & -0.113 & 
~0.031 & 0.261 & -0.669 \\ 
0.9 & -0.606 & 1.292 & -0.221 & -0.719 & -0.290 & 0.494 & -0.729 & -0.117 & 
~0.057 & 0.250 & -0.634 \\ 
1.0 & -0.493 & 1.076 & -0.308 & -0.802 & -0.417 & 0.596 & -0.793 & -0.091 & 
~0.095 & 0.396 & -0.748 \\ \hline
Average & -0.471 & 1.097 & -0.248 & -0.787 & -0.424 & 0.496 & -0.703 & -0.163
& 0.085 & 0.346 & -0.815 \\ 
St. Dev. & ~0.084 & 0.208 & ~0.148 & ~0.107 & ~0.097 & 0.088 & ~0.051 & 
~0.091 & 0.044 & 0.060 & ~0.106 \\
\hline
\end{tabular}
\end{center}
\caption{Selected ESP atomic charges of I$\cdot$SP in the presence of the
protein electrostatic potential.}

\end{sidewaystable}

\begin{sidewaystable}
\centerline{ {\bf TABLE 3.  Intermediate -- and substrate -- Q192G193
peptide unit interactions }}
\begin{center}
\begin{tabular}[!H]{rcc}
& ESP & B.E.  \\ 
\hline

$\Delta $E (I$\cdot$SP) (Complex {\bf I}) & -12(4) & --- \\

$\Delta $E (I$\cdot$SP) (Complex {\bf III}) & -2.6 & --- \\

$\Delta $E (S$\cdot$SP) (Complex {\bf II}) & -6(2) & -4.2 \\

$\Delta $E (S$\cdot$SP) (Complex {\bf IV}) & --2.6 & -1.5 \\

\hline
\end{tabular}
\end{center}
\caption{Energies (kcal/mol) are calculated from the electrostatic
ESP--based model and from binding energies (see Computational Section).
ESP--based energies of complexes {\bf I} and {\bf II} are calculated
as average during the dynamics, whereas those of complexes {\bf III}
and {\bf IV} from the initial structural model.}
\end{sidewaystable}

\begin{sidewaystable}
\centerline{ {\bf TABLE 4.  S$\cdot$SP {\it in vacuo}}}
\begin{center}
\begin{tabular}[!H]{rccccccccccc}
Time~(ps) & O$_\gamma$(S195) & C$_S$ & N$_S$ & O$_S$ & N~(G193) & C~(Q192)
& O~(Q192) & N$_{\delta 1}$(H57) & N$_{\epsilon 2}$(H57) & H$_{\delta 1}$(H57) & O$_{\delta 2}$~(D102)
\\ \hline
Init. & -0.468 & 0.635 & -0.352 & -0.469 & -0.561 & 0.585 & -0.637 & ~0.044
& -0.237 & 0.258 & -0.869 \\ 
0.1 & -0.473 & 0.739 & -0.527 & -0.559 & -0.709 & 0.624 & -0.644 & ~0.135 & 
-0.237 & 0.206 & -0.785 \\ 
0.2 & -0.410 & 0.578 & -0.259 & -0.574 & -0.595 & 0.599 & -0.635 & -0.072 & 
-0.219 & 0.313 & -0.869 \\ 
0.3 & -0.528 & 0.684 & -0.280 & -0.595 & -0.638 & 0.605 & -0.640 & -0.068 & 
-0.458 & 0.317 & -0.879 \\ 
0.4 & -0.476 & 0.585 & -0.219 & -0.519 & -0.642 & 0.633 & -0.643 & -0.156 & 
-0.472 & 0.374 & -0.875 \\ 
0.5 & -0.491 & 0.618 & -0.165 & -0.577 & -0.592 & 0.630 & -0.658 & ~0.070 & 
-0.476 & 0.254 & -0.835 \\ 
0.6 & -0.355 & 0.576 & -0.189 & -0.549 & -0.628 & 0.616 & -0.644 & ~0.149 & 
-0.231 & 0.233 & -0.824 \\ \hline
Average & -0.457 & 0.631 & -0.284 & -0.549 & -0.624 & 0.613 & -0.643 & ~0.015
& -0.333 & 0.279 & -0.848 \\ 
St. Dev. & ~0.053 & 0.057 & ~0.114 & ~0.040 & ~0.044 & 0.016 & ~0.007 & 
~0.107 & ~0.118 & 0.054 & ~0.032 \\
\hline
\end{tabular}
\end{center}
\caption{Selected ESP atomic charges of S$\cdot$SP {\it in vacuo}.}

\end{sidewaystable}

\begin{sidewaystable}
\centerline{ {\bf TABLE 5.  S$\cdot$SP with protein external field}}
\begin{center}
\begin{tabular}[!H]{rccccccccccc}
Time~(ps) & O$_\gamma$(S195) & C$_S$ & N$_S$ & O$_S$ & N~(G193) & C~(Q192)
& O~(Q192) & N$_{\delta 1}$(H57) & N$_{\epsilon 2}$(H57) & H$_{\delta 1}$(H57) & O$_{\delta 2}$(D102)
\\ \hline
Init. & -0.331 & 0.693 & -0.474 & -0.488 & -0.614 & 0.686 & -0.645 & -0.004
& -0.124 & 0.287 & -0.912 \\ 
0.1 & -0.418 & 0.703 & -0.602 & -0.552 & -0.701 & 0.745 & -0.782 & -0.199 & 
-0.256 & 0.345 & -0.803 \\ 
0.2 & -0.349 & 0.492 & -0.389 & -0.513 & -0.574 & 0.722 & -0.768 & -0.217 & 
-0.315 & 0.334 & -0.840 \\ 
0.3 & -0.477 & 0.601 & -0.356 & -0.546 & -0.667 & 0.825 & -0.789 & -0.104 & 
-0.553 & 0.296 & -0.860 \\ 
0.4 & -0.475 & 0.561 & -0.301 & -0.495 & -0.650 & 0.812 & -0.765 & -0.186 & 
-0.576 & 0.343 & -0.843 \\ 
0.5 & -0.455 & 0.553 & -0.262 & -0.524 & -0.570 & 0.772 & -0.787 & -0.129 & 
-0.612 & 0.296 & -0.827 \\ 
0.6 & -0.341 & 0.496 & -0.273 & -0.497 & -0.650 & 0.775 & -0.769 & -0.172 & 
-0.306 & 0.348 & -0.835 \\ \hline
Average & -0.407 & 0.586 & -0.380 & -0.516 & -0.632 & 0.762 & -0.758 & -0.144
& -0.392 & 0.321 & -0.846 \\ 
St. Dev. & ~0.060 & 0.079 & ~0.114 & ~0.023 & ~0.045 & 0.045 & ~0.047 & 
~0.068 & ~0.174 & 0.025 & 0.025 \\
\hline
\end{tabular}
\end{center}
\caption{Selected ESP atomic charges of S$\cdot$SP in the presence of the
protein electrostatic potential.}

\end{sidewaystable}

\end{document}